\documentclass[%
 reprint,
 amsmath,amssymb,
 aps,
 prb,
 floatfix,
]{revtex4-1}

\usepackage{graphicx}
\usepackage{dcolumn}
\usepackage{bm}

\begin{document}

\preprint{APS/123-QED}

\title{Reduced Scaling Hilbert Space Variational Monte Carlo}

\author{Haochuan Wei$^1$}

\author{Eric Neuscamman$^{1,2,}$}%
\email{eneuscamman@berkeley.edu}

\affiliation{
${}^1$Department of Chemistry, University of California, Berkeley, CA, 94720, USA \\
${}^2$Chemical Sciences Division, Lawrence Berkeley National Laboratory, Berkeley, CA, 94720, USA
}

\date{\today}

\begin{abstract}
We show that for both single-Slater-Jastrow and Jastrow geminal power wave functions, the formal cost scaling of
Hilbert space variational Monte Carlo can be reduced from fifth to fourth order in the system size,
thus bringing it in line with the long-standing scaling of its real space counterpart.
While traditional quantum chemistry methods can reduce costs related to the two-electron integral tensor through
resolution of the identity and Cholesky decomposition approaches, we show that such approaches are ineffective in the
presence of Hilbert space Jastrow factors.
Instead, we develop a simple semi-stochastic approach that can take similar advantage of the
near-sparsity of this four-index tensor.
Through demonstrations on alkanes of increasing length, we show that accuracy and overall statistical uncertainty
are not meaningfully affected and that a total cost crossover is reached as early as 50 electrons.
\end{abstract}

\maketitle

\section{Introduction}
\label{sec:intro}

Quantum Monte Carlo (QMC) methods that rely on random walks in the space of Slater determinants, i.e.\ in Hilbert space (HS),
have seen rapid progress in recent years.
Work on full configuration interaction QMC (FCIQMC) \cite{booth2009fermion,petruzielo2012semistochastic}
has laid bare how effectively sparsity in a wave function's
determinant expansion can be exploited, helping for example to reignite interest in selective configuration interaction methods
\cite{Holmes:2016:heatBath,caffarel2016communication,Tubman:2016:asci,Umrigar:2017:semistoc_heat_bath_CI}
as general-purpose tools for strongly correlated systems.
Likewise, auxiliary field QMC has seen rapid progress in its efficacy in molecules,
\cite{purwanto2008eliminating,purwanto2009excited,morales:2018:nociafqmc}
including some that display strong electron correlation. \cite{purwanto2015auxiliary}
In addition to these developments in HS projector Monte Carlo,
progress in Hilbert space variational Monte Carlo (HSVMC) has provided an exact size-consistency correction \cite{Eric:2012:prl_jagp}
for the antisymmetric geminal power (AGP) ansatz and has facilitated more general efforts in wave function stenciling.
\cite{Eric:2016:ScO,neuscamman2016subtractive,Brett:2017:num_count_jast}
Indeed, work in stochastic HS methods remains very active, with efforts towards a post-HSVMC perturbation theory appearing earlier this year.
\cite{Sharma:2018:stochpt}

Although they have significant differences from one another, these HS approaches enjoy a number of advantages in common
when compared to QMC approaches that rely on random walks in real space.
For one, they can interoperate with and be compared to traditional quantum chemistry methods more easily, as they live within the same
basis set approximation and provide relatively easy access to standard wave function characteristics such as natural orbitals
and reduced density matrices. \cite{overy2014unbiased}
Furthermore, like essentially all HS methods, they can employ the frozen core approximation,
\cite{Aikens:2003:frozen_orbital,Kiejna:2006:frozen_core,Patkowski:2007:frozen_core}
which unlike pseudopotentials provides a rigorous accounting of Pauli exclusion effects between core and valence electrons.
As the construction of accurate and transferable pseudopotentials for transition metals remains a challenging and active area
of research \cite{trail2015correlated} and is particularly difficult in the context of real space QMC, \cite{dzubak2017quantitative}
HS methods offer a potentially powerful advantage in this broad area of chemistry.
Finally, the unfavorable cost scaling of all-electron real space QMC with nuclear charge \cite{Ceperley:1986:DMC,Hammond:1987:valence_qmc} poses challenges for core excitation spectroscopy, whose ability to probe local electronic structure in an increasingly time
resolved manner \cite{loh2013capturing} promises to enhance its already considerable role in chemical spectroscopy.
With recent progress in excited state VMC \cite{Chris:2016:omega,Jacki:2017:omega,blunt2017charge} promising to greatly
enhance simple wave functions' capture of the orbital relaxation effects that are especially important in core spectroscopy,
\cite{Besley:2009:delta_scf} HSVMC is in principle well-placed to assist in the design and interpretation of such experiments.

However, HSVMC faces the significant disadvantage of having to deal directly with the four-index electron-electron repulsion tensor within
the HS representation of the ab initio Hamiltonian.
While trivial in real space QMC, these electron-electron repulsion terms often drive the cost scaling of HSVMC, as they typically cause any
randomly sampled configuration to connect to $O(n^4)$ configurations within the wave function.
As random walks are typically performed in a local basis in order to maximize the efficacy of the Jastrow factor,
\cite{Eric:2013:jcp_jagp,Neuscamman:2013:or_jagp} this issue exists even
when the fermionic part of the wave function is the Hartree-Fock determinant, as its representation in local orbitals is dense.
When combined with the need to increase sample size linearly with system size to control statistical uncertainty, one finds that
the energy evaluation for even the simplest Jastrow-Slater wave function requires $O(n^5)$ operations in HSVMC, even though the
analogous wave function's evaluation in real space requires only $O(n^4)$ operations. \cite{Foulkes:2001:QMC_solids}
While one might expect relief to come from standard quantum chemistry approaches such as the resolution of the identity (RI) \cite{Weigend:1997:RI_MP2,Neese:2003:resolution_identity,Ren:2012:res_identity}
or Cholesky decomposition (CD), \cite{Gentle:1998:Cholesky,Nash:1990:Cholesky,koch2003reduced}
the HS Jastrow factor connects all four of the repulsion tensor's indices in a way that prevents these rank-reduction
methods from offering any advantage.
There is thus a disparity between the cost scaling of VMC in real space and Hilbert space that cannot be immediately remedied by
standard methods, despite the fact that these two QMC techniques are working with wave function forms that are the analogues of each other.

To resolve this disparity and bring the cost scaling of HSVMC in line with that of its real space cousin, we will employ a semi-stochastic
summation of the two-electron integral terms in which the large terms, which are small in number, are summed exactly while the large sum over
small terms is estimated statistically.
This type of semi-stochastic approach has proven very effective in both FCIQMC, \cite{petruzielo2012semistochastic} where it has been used
to perform exact imaginary time propagation among the ``important'' configurations in conjunction with stochastic propagation among the rest,
and in the perturbative correction to heat bath configuration interaction, \cite{Umrigar:2017:semistoc_heat_bath_CI}
where the summation of second order terms is broken into a small set of large terms and a large set of small terms.
As HS Jastrow factors work best in a local orbital basis, \cite{Eric:2013:jcp_jagp,Neuscamman:2013:or_jagp} the vast majority of two-electron integrals that HSVMC encounters are small in magnitude, and so the sum over two-electron integrals that drives the method's cost scaling is very
well-suited to a semi-stochastic treatment.
While the eventual sparsity of the integral tensor in large systems provides a formal guarantee that simply exploiting sparsity would
reduce cost scaling in the large-system limit, our semi-stochastic approach becomes advantageous already for about 50 electrons due to
its ability to exploit the presence of many small but not-quite-negligible integrals.
In this proof-of-principle study, we will describe why traditional approaches are ineffective, the details of our semi-stochastic approach,
and benchmarking calculations on alkanes that numerically confirm our formal conclusion that the disparity in cost scaling
between real space and Hilbert space VMC can be eliminated.

\section{Theory}
\label{sec:theory}

\subsection{Hilbert-space VMC}
\label{sec::tsub1}

We begin with a generic total energy expression given a Hamiltonian $H$ and an electronic state $\Psi$.
\begin{align}
E = \frac{ \langle \Psi | H | \Psi \rangle } {\langle \Psi | \Psi \rangle}
\end{align}
Expressing the Slater determinant basis of our HS using occupation number (ON) vectors \cite{Helgaker:2000} $|\vec{n}\rangle$, we can resolve an identity and write the total energy as sum of ``local'' energies $E_L(\vec{n})$ weighted by the wave function's probability distribution.
\begin{align}
\label{eqn:energy}
E &= \frac{ \sum_{\vec{n}}  \langle \Psi | \vec{n} \rangle \langle \vec{n} |H | \Psi \rangle } {\langle \Psi | \Psi \rangle}
  = \sum_{\vec{n}} P(\vec{n}) E_{L}(\vec{n}) \\
  & \qquad \qquad \hspace{1.8mm} P(\vec{n}) \equiv \frac{ | \langle \Psi | \vec{n} \rangle |^{2}}{\langle \Psi | \Psi \rangle} \\
& \qquad \qquad E_L(\vec{n}) \equiv \frac{\langle \vec{n} |H | \Psi \rangle}{\langle \vec{n} | \Psi \rangle}
\end{align}
VMC achieves an unbiased statistical estimate of this total energy
\begin{align}
E_{{}_{\mathrm{VMC}}} = \frac{1}{N} \sum_{\vec{m} \in S} E_{L}(\vec{m})
\end{align}
by averaging the local energies on a sample $S$ that consists of $N$ configurations drawn from from $P$ via the Metropolis algorithm. \cite{metropolis1953}
Although other choices of the importance sampling function are possible and often advantageous, \cite{Trail:2008:heavy_tail,Trail:2008:alt_sampling,robinson2017} for the present study the simple choice of the wave function probability distribution $P$ will suffice.
Using the Central Limit Theorem, \cite{Spiegel:1992:clt} the statistical uncertainty $\epsilon$ in this energy estimate can be related to the energy variance
\begin{align}
\sigma^2 &= \left\langle (E_L - E )^2 \right\rangle_P \\
&= \sum_{\vec{n}} P(\vec{n}) \left( E_L(\vec{n}) - E \right)^2
\end{align}
as $\epsilon \propto \sigma/\sqrt{N}$.
Although the practical need to take a large number of samples to control this uncertainty is a drawback of Monte Carlo approaches, the sample size required by VMC is smaller than one might expect due to the zero variance principle. \cite{toulouse2016introduction} 

The second-quantized ab initio Hamiltonian 
\begin{align}
H = \sum_{pq} h_{pq} a^+_p a_q + \frac{1}{2} \sum_{pqrs} (pq|rs) a^+_p a^+_r a_s a_q
\end{align}
causes the naive cost of Hilbert-space VMC to scale as at least $O(n^4)$
per configuration sample due to the 4-index sum over two-electron integrals.
This scaling with respect to the electron number $n$ assumes that the number of spatial orbitals is proportional to $n$.
Note that the scaling may be even worse than this depending on the complexity of $\Psi$, but for common choices such as the Slater-Jastrow and Jastrow AGP (JAGP) wave functions the use of the Sherman-Morrison formula and the Woodbury matrix identity allow the cost scaling to be limited to that dictated by the above sum over double excitations. \cite{nukala2009fast,Eric:2013:jcp_jagp}
In contrast, in real-space VMC, where the identity in Eq.\ (\ref{eqn:energy}) is resolved in the real-space position basis, the cost of evaluating one local energy scales as $O(n^3)$ and is driven by the kinetic energy term rather than the electron-electron repulsion. \cite{Foulkes:2001:QMC_solids}
As the only difference between these two cases is which basis the expressions are being evaluated in, one might imagine that the $O(n^4)$ local energy scaling of HSVMC is artificially high, and indeed the success of various quantum chemistry methods at reducing the cost scaling of the electron-electron repulsion integrals suggests that a similar scaling reduction should be possible in HSVMC.

\subsection{The JAGP ansatz}
\label{sec::jagp}

We will demonstrate our scaling reduction using the JAGP ansatz. \cite{Eric:2013:jcp_jagp}
\begin{align}
&|\Psi\rangle = e^{\hat{J}} |\Psi_{\mathrm{AGP}}\rangle \\
&\hat{J} =   \sum_{p\leq q} J^{\alpha \alpha}_{pq} \hat{n}_p \hat{n}_q
           + \sum_{\bar{p}\leq \bar{q}} J^{\beta \beta}_{\bar{p}\bar{q}} \hat{n}_{\bar{p}}\hat{n}_{\bar{q}}
           + \sum_{p \bar{q}} J^{\alpha \beta}_{p\bar{q}} \hat{n}_p \hat{n}_{\bar{q}} \\
&|\Psi_{\mathrm{AGP}}\rangle = \left( \sum_{r \bar{s}} F_{r \bar{s}} a_r^{\dagger} a_{\bar{s}}^{\dagger} \right)^{n/2} |0\rangle
\end{align}
Here indices without (with) bars are for alpha (beta) spin orbitals, $a^\dagger_p$ creates an electron in the $p$th orbital,
$\hat{n}_p = a^\dagger_p a_p$, and $|0\rangle$ is the vacuum state.
Note that the JAGP has also been used extensively in in real space QMC,
\cite{Sorella:2003:jagp,Sorella:2009:jagp_molec,sorella:2011:h_chain,Guidoni:2012:ethylene,Coccia2014}
where, like Slater-Jastrow, it enjoys a cost that scales quartically with system size (cubically for intensive quantities).
Note also that, if the symmetric pairing matrix $F$ is chosen to have rank $n/2$, then the AGP wave function simplifies to a single
Slater determinant, \cite{Sorella:2009:jagp_molec} and so the JAGP ansatz contains Slater-Jastrow as a special case.
Thus, if we can demonstrate reduced scaling for JAGP in HS, the same follows for Slater-Jastrow.

\subsection{Resolution of the Identity}
\label{sec::tsub3}

Reduced-scaling methods in quantum chemistry often rely on tensor factorization to simplify the tensor contractions
involved in energy expectation values and other expressions.
One of the most well-known and widely used factorization approaches is to apply an approximate resolution of the identity (RI)
\cite{Weigend:1997:RI_MP2,Neese:2003:resolution_identity,Ren:2012:res_identity}
within the two-electron integral tensor,
\begin{align}
(pq|rs) & = \int \int d\mathbf{r}_1 d\mathbf{r}_2 \frac{ \psi_p(\mathbf{r}_1) \psi_q(\mathbf{r}_1)
                                                         \psi_r(\mathbf{r}_2) \psi_s(\mathbf{r}_2) }{ | \mathbf{r}_1 - \mathbf{r}_2 | }
\label{eqn:tei_defn} \\
& \approx \sum_{\sigma \tau} (pq|\sigma) M_{\sigma \tau}^{-1} (\tau|rs)
\end{align}
in which the dimension of the matrix $\mathbf{M}$ is greatly smaller than the square of the number of one-electron basis functions.
Typical choices for obtaining $\mathbf{M}$ and the three-index tensors $(pq|\sigma)$ and $(\tau|rs)$ are the
density fitting \cite{sodt2006linearDF} and Cholesky \cite{weigend2009cholesky_vs_ri} approaches.
In practice, the dimension of $\mathbf{M}$ need grow only linearly with system size in order for the RI approach
to be accurate, which leads to a substantial reduction in memory use and, in many cases, a reduction in the cost scaling of a
method due to simplifications in the method's tensor contractions that the RI decomposition permits.

Unfortunately, not all contractions against the two-electron integrals can benefit from the RI approach.
Consider an example relevant to the present discussion, which is the component of the JAGP local energy coming
from the all-$\alpha$ term in the two-electron part of the Hamiltonian.
If the Jastrow factor is not present, this energy contribution is \cite{Eric:2013:jcp_jagp}
\begin{align}
E^{AGP}_{2 \alpha \alpha} &= \frac{1}{2} \sum_{iajb}(ia|jb) \frac{\langle \vec{n} | a^{\dagger}_{i}a_{a}a^{\dagger}_{j}a_{b} | \Psi_{AGP} \rangle } { \langle \vec{n} | \Psi_{AGP} \rangle } \\
&= \frac{1}{2} \sum_{iajb}(ia|jb) [(R \Theta)_{ia} (R \Theta)_{jb} - (R \Theta)_{ja} (R \Theta)_{ib}]
\label{eqn:le_agp_all_alpha}
\end{align}
where $i$ and $j$ run over the $n_o$ occupied $\alpha$ orbitals in the sampled configuration $\vec{n}$,
$a$ and $b$ run over the $n_u$ unoccupied $\alpha$ orbitals,
and the matrices $R$ and $\Theta$ are derived from the AGP pairing matrix.
The left-hand part of the sum in Eq.\ (\ref{eqn:le_agp_all_alpha}) is a Coulomb-like term whose cost of
evaluation can be lowered by exploiting the RI approach to the integrals
\begin{align}
\notag
&  \frac{1}{2} \sum_{iajb}(ia|jb) (R \Theta)_{ia} (R \Theta)_{jb} \\
&\simeq \frac{1}{2} \sum_{iajb} \sum_{\sigma \tau} (R \Theta)_{ia} (ia|\sigma) M_{\sigma\tau} (\tau|jb) (R \Theta)_{jb}
\end{align}
so that the worst-case step in the contraction is the $ia$ (or $jb$) sum with cubic system-size scaling.
The right-hand part of the sum in Eq.\ (\ref{eqn:le_agp_all_alpha}) is exchange-like, and although such terms are not as easy
to simplify with RI approaches, they can in many cases be handled at a dramatically lower cost than the naive direct sum would suggest.
\cite{Goedecker:1999:linear_elec_struc,Parr:1995:dft_review,Lin:2013:pexsi_ks_dft}

With the inclusion of the Jastrow factor, however, this part of the local energy expression is modified by what can be
written as a transformation of the two-electron integral tensor \cite{Eric:2013:jcp_jagp}
\begin{align}
\notag
(ia|jb) \rightarrow (ia|jb) \hspace{0.5mm} \mathrm{exp}\big(
& \hspace{2.0mm} K_a + K_b - K_i - K_j \\
\notag
& \hspace{2.0mm} + J_{ab} - J_{ia} - J_{jb} \\
& \hspace{2.0mm} + J_{ij} - J_{ib} - J_{ja}
  \hspace{5.0mm} \big)
\end{align}
in which $K$ depends both on $\vec{n}$ and the Jastrow variables.
This transformation spoils the simple RI factorization of the coulomb-like term
due to the links created between pairs of two-electron integral indices,
which also serve to greatly complicate the handling of the exchange-like term.
We therefore see that JAGP, due to the effects of the Jastrow factor,
is not easily accelerated by the application of RI techniques, and so we must look
elsewhere if we wish to lower the cost-scaling of HSVMC to match that of real space VMC.

\subsection{Semi-stochastic Local Energies}
\label{sec::ssle}

\begin{figure}[b]
  \includegraphics[width=8.0cm,angle=0]{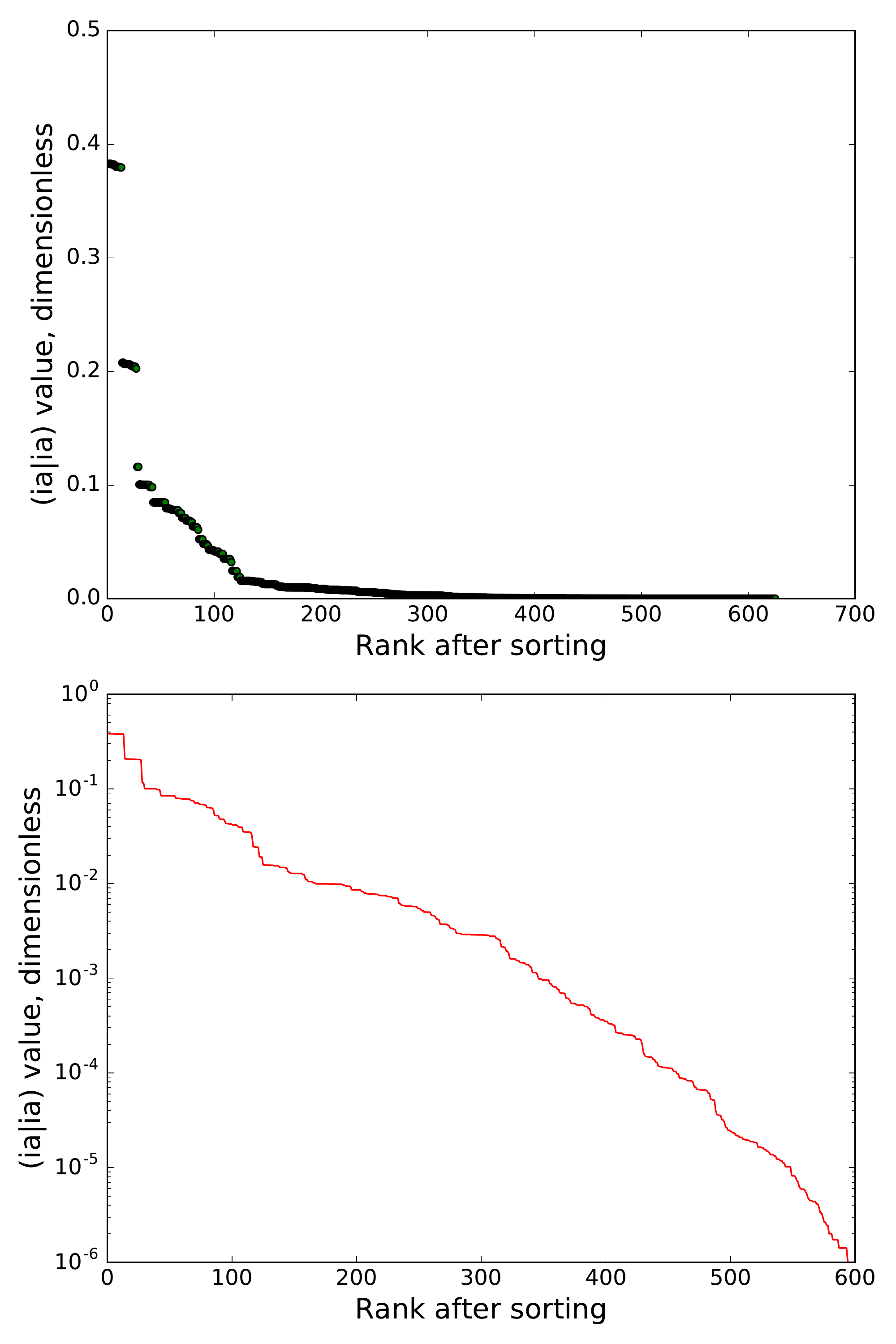}
  \caption{Sorted integral values for different index pairs ($i$,$a$) used in Cauchy-Schwartz bounds,
           arranged on both regular (top) and logarithmic (bottom) scales for octane in a symmetrically orthonormalized STO-3G basis.
  }
  \label{fig:causch_integral_decay}
\end{figure}

Rather than employing tensor factorization, the local energy can be evaluated efficiently by
breaking the two-electron sum over $iajb$ into a small sum of large terms and a large sum of small terms,
the latter of which is estimated statistically.
We rely on the Cauchy-Schwartz inequality \cite{Arfken:1985:cauchy_schwarz}
\begin{align}
(ia|jb) \leq |(ia|jb)| \leq \sqrt{|(ia|ia)|} \sqrt{|(jb|jb)|}
\end{align}
to decide which terms to place in which sum.
As shown for example in Figure \ref{fig:causch_integral_decay}, these bounds imply that most integrals
are small in value when working in a local basis.
In the limit of a large system size, local one-particle bases (which are anyways most appropriate for encoding correlations via the Jastrow)
lead this inequality to imply that the number of non-trivial elements in the two-electron integral tensor will
scale only quadratically with system size.
Indeed, for a term to be non-trivial, orbital $a$ must be among the $O(1)$ orbitals that have
overlap with orbital $i$.
Likewise for orbitals $j$ and $b$.
As we work in the local basis obtained via symmetric orthogonalization \cite{Szabo:symm_ortho}
of the atomic orbitals, we may thus use Cauchy-Schwartz as a guide for identifying
large groups of small integrals that can be handled stochastically without incurring
substantial increases in statistical uncertainty.

The precise computational cost of the semi-stochastic evaluation consists of three parts:
screening the terms into deterministic and stochastic portions,
computing the deterministic portion, and computing the stochastic portion.
Calculating or looking up the 2-index $(ia|ia)$ and $(jb|jb)$ integrals requires $O(n^2)$ operations,
and sorting those integrals requires $O(n^2 \mathrm{log}(n))$ operations. \cite{Sedgewick:1977:quicksort}
Hence, the cost of screening is $O(n^2 \mathrm{log}(n))$.
\begin{figure}[b]
  \includegraphics[width=8.0cm,angle=0]{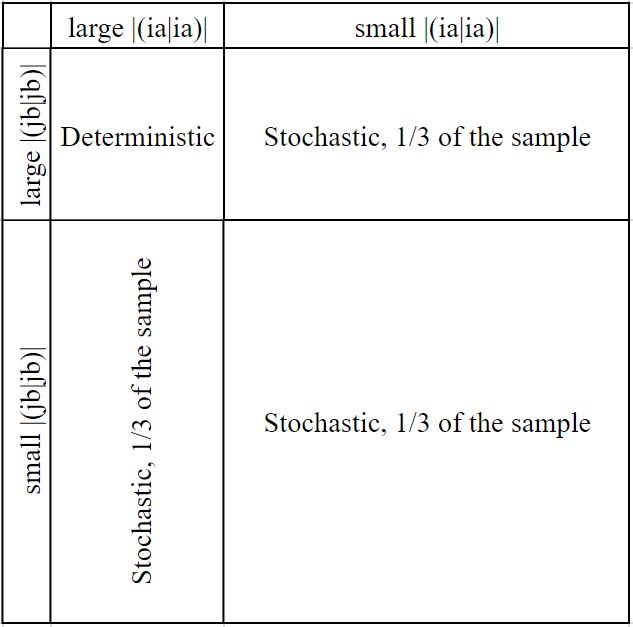}
  \caption{A schematic showing the division of two-electron integrals $(ia|jb)$ into deterministic and stochastic bins
           according to the size of the Cauchy-Schwartz bounds for each index pair.
  }
  \label{fig:block_illustration}
\end{figure}
After screening, we may divide integrals into four groups, as shown in Figure \ref{fig:block_illustration}.
First, we find the set of $L$ index pairs ($i$,$a$) with the largest $|(ia|ia)|$ values 
and deterministically add up the two-electron parts of the local energy that involve two-electron integrals
for which both the ($i$,$a$) and ($j$,$b$) index pairs are within this set.
To achieve the desired overall cost scaling, we choose the number of ``large'' index pairs as $L\sim O(n^{3/2})$,
which also ensures that in the large system limit only trivial terms will be left over for the stochastic part.
In the benchmarks below, we have made the specific choice of $L = \mathrm{floor}(\sqrt{n_o n_u^2})$.
Finally, for the summation over the ``small'' terms in the other three groups in Figure \ref{fig:block_illustration},
we make a statistical estimate by drawing index quadruplets from a uniform distribution over those quadruplets in each group.
Again, to ensure that the overall scaling with system size matches that of real space VMC, we set the size of the
sample (divided evenly across the three groups) as $O(n^2)$.
In the benchmarks below, we use the specific sample size of $8n^2$.
Note that this statistical estimation affects only the two-electron part of the local energy; all one-electron terms are
evaluated deterministically as has been implemented previously. \cite{Eric:2013:jcp_jagp}

We can also analytically estimate the statistical uncertainty of the total energy estimate. If we compute the local energy at least partially stochastically, then the total energy estimate can be written as
\begin{align}
\frac{1}{N} \sum_{\vec{m}\in S} E_{L}(\vec{m}) \simeq \frac{1}{N} \sum_{\vec{m}\in S} \tilde{E}_{L}(\vec{m})
\end{align}
where $\tilde{E}_{L}(\vec{m})$ denotes a stochastic estimate of the local energy.
Let us denote by a random variable $\delta_{\vec{m}}$ the difference between the semi-stochastic and deterministic local energy values
\begin{align}
\delta_{\vec{m}} = \tilde{E}_{L}(\vec{m}) - E_{L}(\vec{m}).
\end{align}
for each of the configurations $\vec{m}$ that appear in our VMC sample $S$.
\begin{figure}[b]
  \includegraphics[width=8.0cm,angle=0]{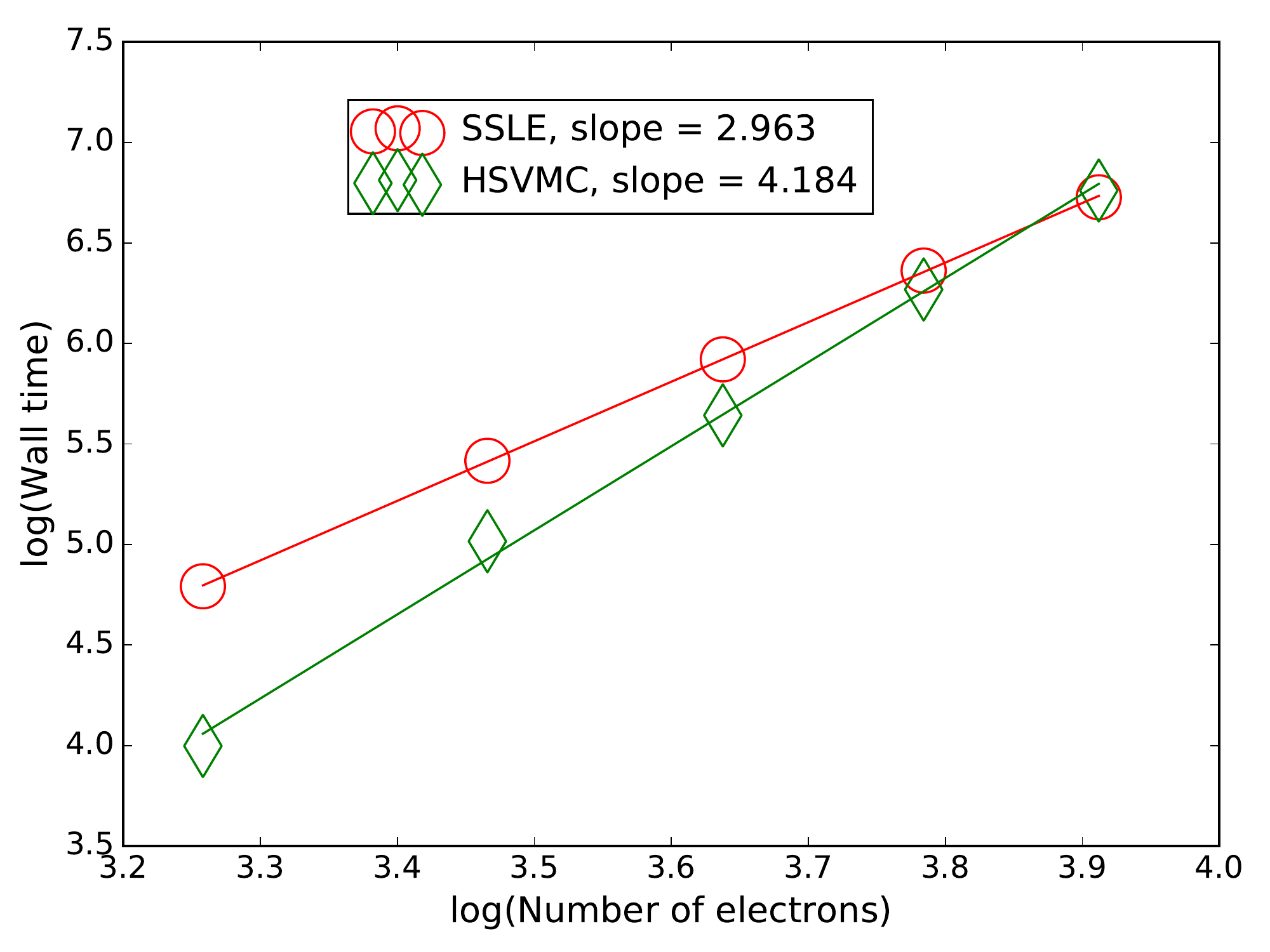}
  \caption{Scaling of the portion of the calculation that computes the local energy and its derivative with regard to the wave function parameters. The number of configuration samples is fixed. The SSLE approach shows an $O(n^3)$ scaling, whereas the traditional, deterministic local energy approach shows a $O(n^4)$ scaling.
  }
  \label{fig:scaling_ledr}
\end{figure}
As we are using independent, uniform random samples of two-electron integral index quadruplets for our stochastic local energy estimates at each
of the $N$ configuration samples, these random variables will be uncorrelated, and so upon summing our local energy estimates to generate our total energy
estimate the variances of these random variables will be additive. \cite{Loeve:1977:Probability}
This allows us to express the variance in our total energy estimate as
\begin{align}
\sigma^2 &= \langle (E_{L}(\vec{m}) - E )^2 \rangle_{|\Psi|^{2}} + \frac{1}{N} \sum_{\vec{m}\in S} \mathrm{var}\left(\delta_{\vec{m}}\right)
\label{eqn:variance_analysis}
\end{align}
where $\mathrm{var}(\delta_{\vec{m}})$ is the variance in our local energy estimate for configuration $\vec{m}$.
The first term here is the usual VMC energy variance due to the random sample of configurations, whereas the second term is the additional energy variance due to
the use of semi-stochastic local energy (SSLE) estimates.
Crucially, we see that the statistical uncertainty in the total energy, which is proportional to $\sqrt{\sigma^2/N}$, will still decrease
at the same $1/\sqrt{N}$ rate as in typical VMC, although its magnitude at a given fixed $N$ will of course depend on the size of the sample we use for the stochastic portion of the local energy.

\section{Results}
\label{sec:results}

\subsection{Scaling Reduction}
\label{sec::rsub1}

\begin{figure}[b]
  \includegraphics[width=8.0cm,angle=0]{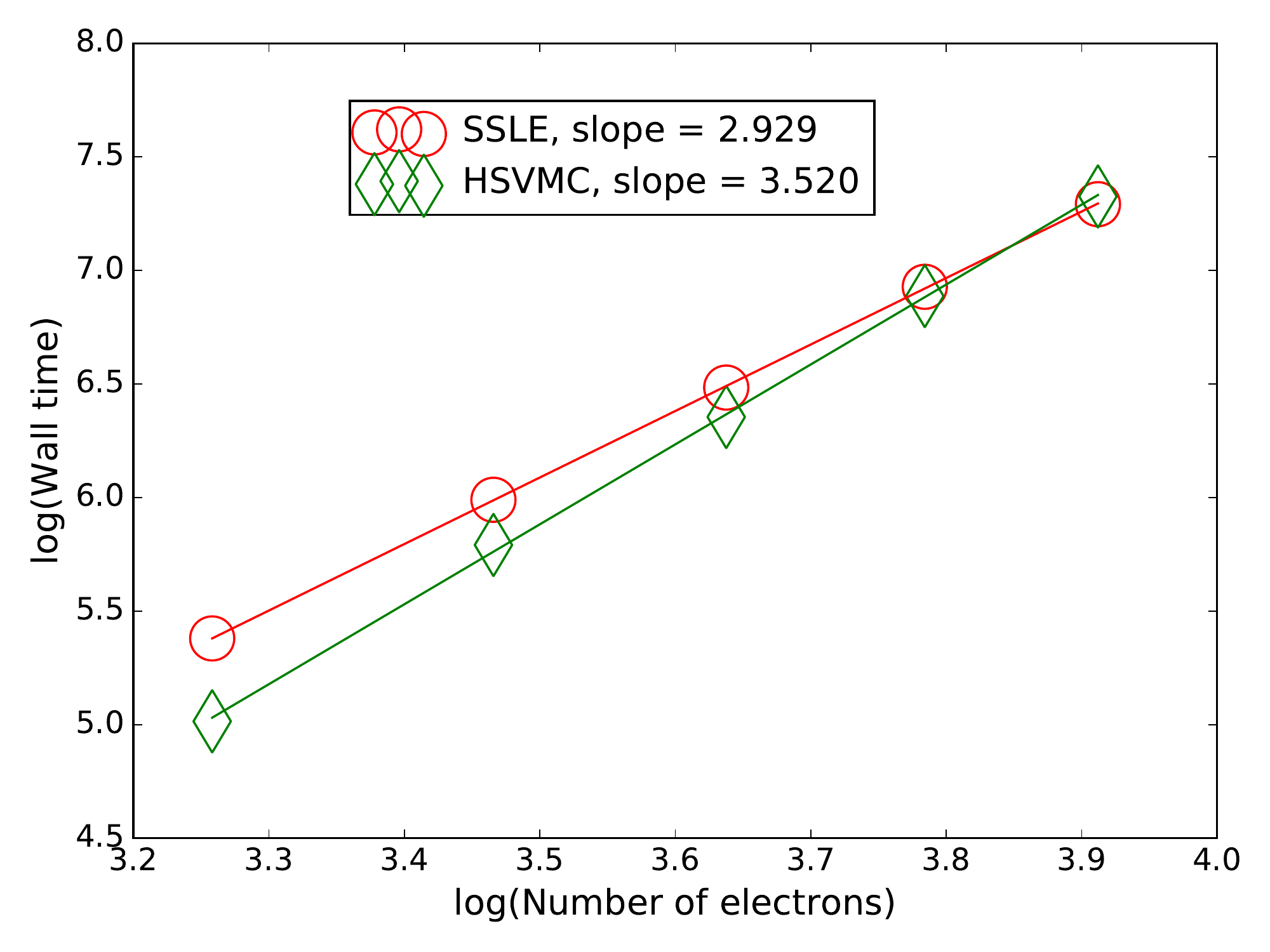}
  \caption{Scaling of the full total energy calculation. The number of configuration samples is fixed. The SSLE approach shows an $O(n^3)$ scaling, whereas the traditional, deterministic local energy approach shows an overall scaling between $O(n^3)$ and $O(n^4)$. This matches our expectation, because the Markov chain propagation scales as $O(n^3)$.
  }
  \label{fig:scaling_fqmc}
\end{figure}

We have verified the theoretically expected scaling reduction for our SSLE approach in a series of linear alkanes,
from butane to octane, treated in the STO-3G orbital basis and using the Hartree-Fock ground-state geometries from the
Computational Chemistry Comparison and Benchmark Database. \cite{NIST:2016:geoms}
As seen in Figures \ref{fig:scaling_ledr} and \ref{fig:scaling_fqmc} for a straightforward evaluation of the JAGP energy,
SSLE brings the cost scaling per configuration sample down to cubic in the system size, both for the cost of the local energy evaluations
alone and for the overall cost of the entire HSVMC energy estimate, which also includes propagation of the Metropolis Markov chain.
Note that the number of electrons in these plots excludes those in the frozen core, which we took to be all the 1s electrons on
the carbons.
As expected, the cost of the local energy evaluations in traditional HSVMC scales quartically with system size due to
sums over the four-index two-electron integrals.
The observed scaling of the overall energy evaluation, including Markov chain propagation, is between cubic and quartic, which can be
understood by recognizing that the Markov chain propagation step has a cubic theoretical asymptotic scaling, and so in these
system sizes a scaling in between that of the asymptotic scalings of the chain propagation and the local energy evaluation
is observed.
Of particular note is the fact that for this class of systems, we achieve not only a scaling reduction but also a
cost crossover in between heptane and octane, i.e.\ at about 50 valence electrons.

\subsection{Statistical Uncertainty}

\begin{figure}[t]
  \includegraphics[width=8.0cm,angle=0]{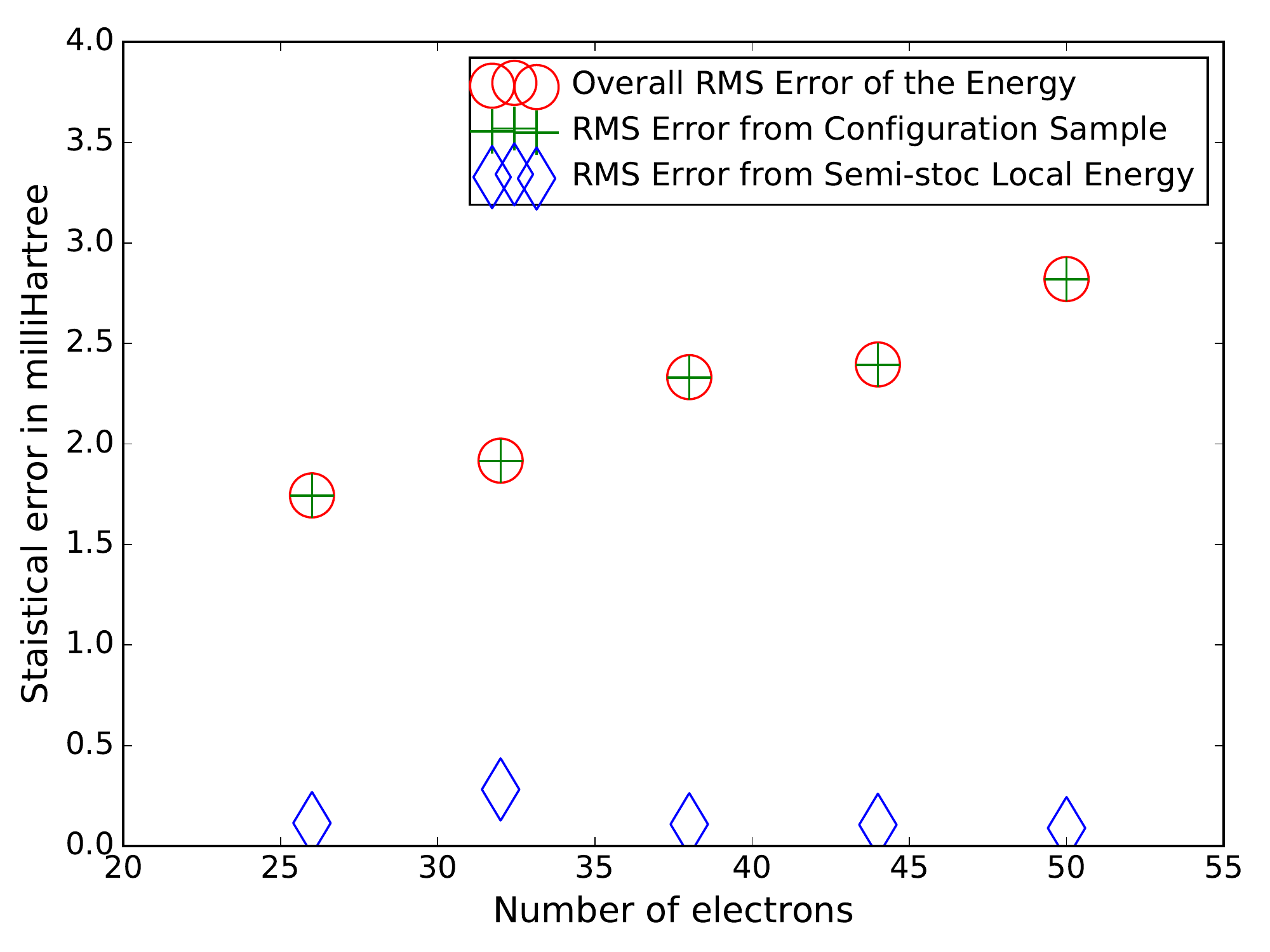}
  \caption{Statistical error of the total energy for the calculations in Figures \ref{fig:scaling_ledr} and
           \ref{fig:scaling_fqmc}.
           The overall statistical uncertainty (circles)
           and the uncertainty from the configuration sampling (crosses) are obtained from blocking analyses.
           The uncertainty due to the SSLE itself (diamonds) is estimated by repeated evaluations
           on the same sample of configurations, each with independent index quadruplet samples.
  }
  \label{fig:accuracy_vary_system}
\end{figure}

\begin{figure}[b]
  \includegraphics[width=8.0cm,angle=0]{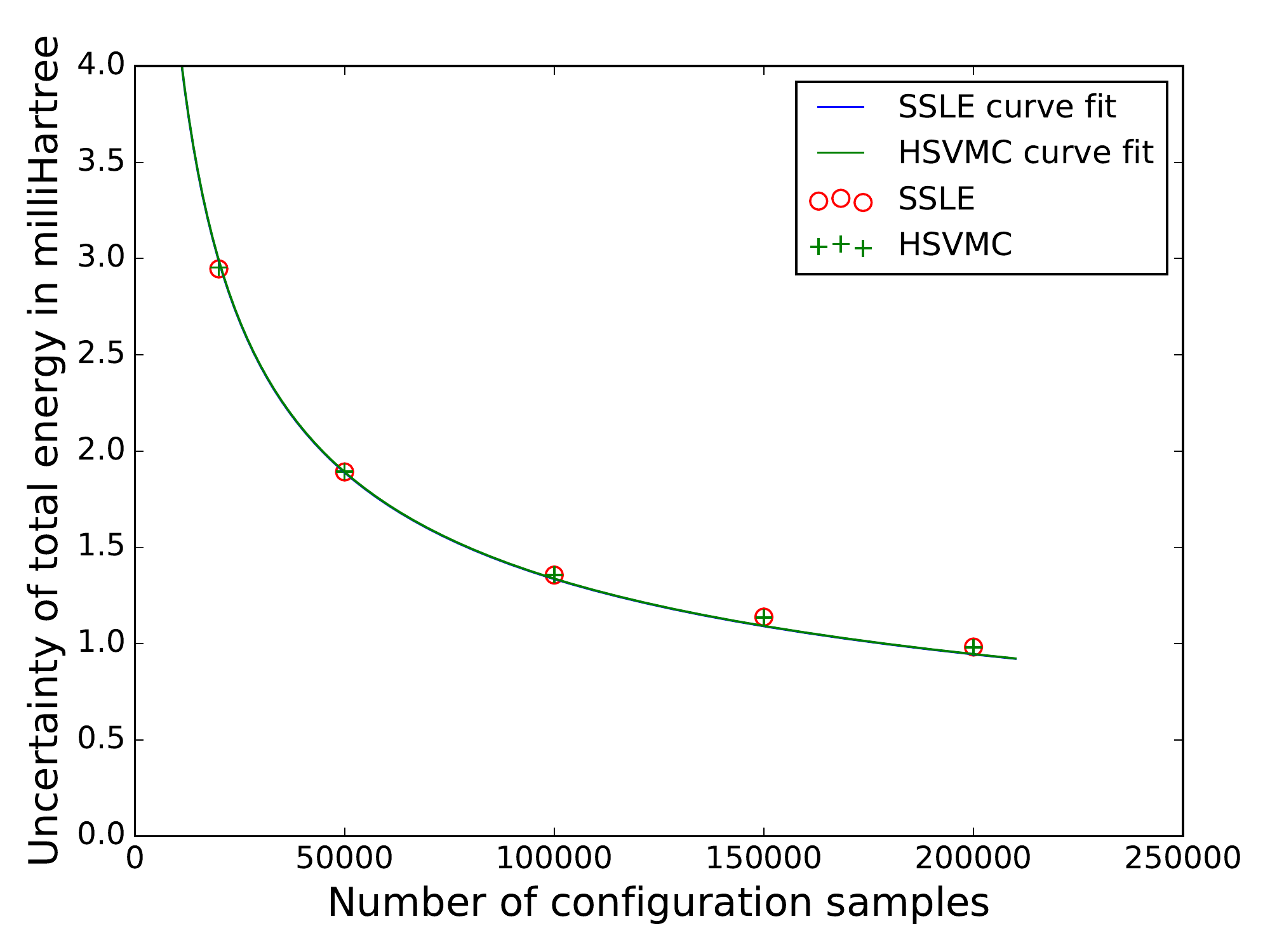}
  \caption{Statistical uncertainty of the total energy vs. the number of configuration samples. The system is STO-3G pentane. The fitted curve, whose expression is $y = c \sqrt{\frac{1}{x}}$ where c is a parameter, indicates that the uncertainty decays as approximately $\sqrt{\frac{1}{N}}$ both with and without the use of SSLE.
  }
  \label{fig:accuracy_vary_vmc_pe}
\end{figure}

Reducing the scaling and cost of HSVMC would not be worthwhile if it came at the cost of large
increases in statistical uncertainty.
In Figure \ref{fig:accuracy_vary_system}, we see that although the SSLE approach does introduce
an additional component to the uncertainty as we would expect from Eq.\ (\ref{eqn:variance_analysis}),
it is small and decreases in importance as the system size grows.
As these calculations were carried out with a system-independent value of the configuration sample size $N$
in order to reveal the per-configuration cost scaling, we see that this decay in the added SSLE uncertainty
with system size is not due to configuration sampling.
Instead, this comes from the fact that, asymptotically, there are only $O(n^2)$ non-trivial two-electron
integral elements while we have designed the SSLE to handle $O(n^3)$ elements deterministically.
Thus, as system size grows, the size of the two-electron integrals that are treated stochastically should
decay to zero, thus driving down the SSLE's contribution to statistical uncertainty in larger systems.
Furthermore, Eq.\ (\ref{eqn:variance_analysis}) teaches us that this benefit will remain even when we
increase the number of configuration samples $N$, and so in a large system where $N$ must grow linearly
with system size to control configuration sampling uncertainty in the total energy, the absolute energetic
uncertainty coming from the SSLE approach will decay even faster.

As an additional verification that the SSLE does not modify the usual statistical properties of HSVMC,
we have checked explicitly that the central limit theorem's $1/\sqrt{N}$ reduction in overall uncertainty
is maintained.
As seen in Figure \ref{fig:accuracy_vary_vmc_pe}, the expected reduction in uncertainty with increasing
numbers of configuration samples is maintained.
To summarize, we find that in the evaluation of the JAGP energy, the SSLE approach does not meaningfully
affect the statistics of HSVMC and in fact even the small effect it does have decays with system size.

\subsection{Optimizing the Wave Function}

While reducing the cost and scaling of the energy evaluation in HSVMC is welcome, it is not of much
practical use if the benefits do not extend to the wave function optimization.
This reality is why in our cost scaling data in Figures \ref{fig:scaling_ledr} and \ref{fig:scaling_fqmc}
we have analyzed the time for evaluating the local energy \textit{and} its derivatives with respect to
all wave function variables, as these derivatives are the foundation of optimization methods.
As the cost of estimating these derivatives typically dominate the cost of wave function optimization,
whether working with the linear method \cite{Nightingale:2001:linear_method, Umrigar:2005:lm,Umrigar:2007:newton_lm_perturb}
or an accelerated descent scheme, \cite{schwarz2017projector} it is crucial that their evaluation also
be made more efficient by the SSLE approach.

\begin{figure}[t]
  \includegraphics[width=8.0cm,angle=0]{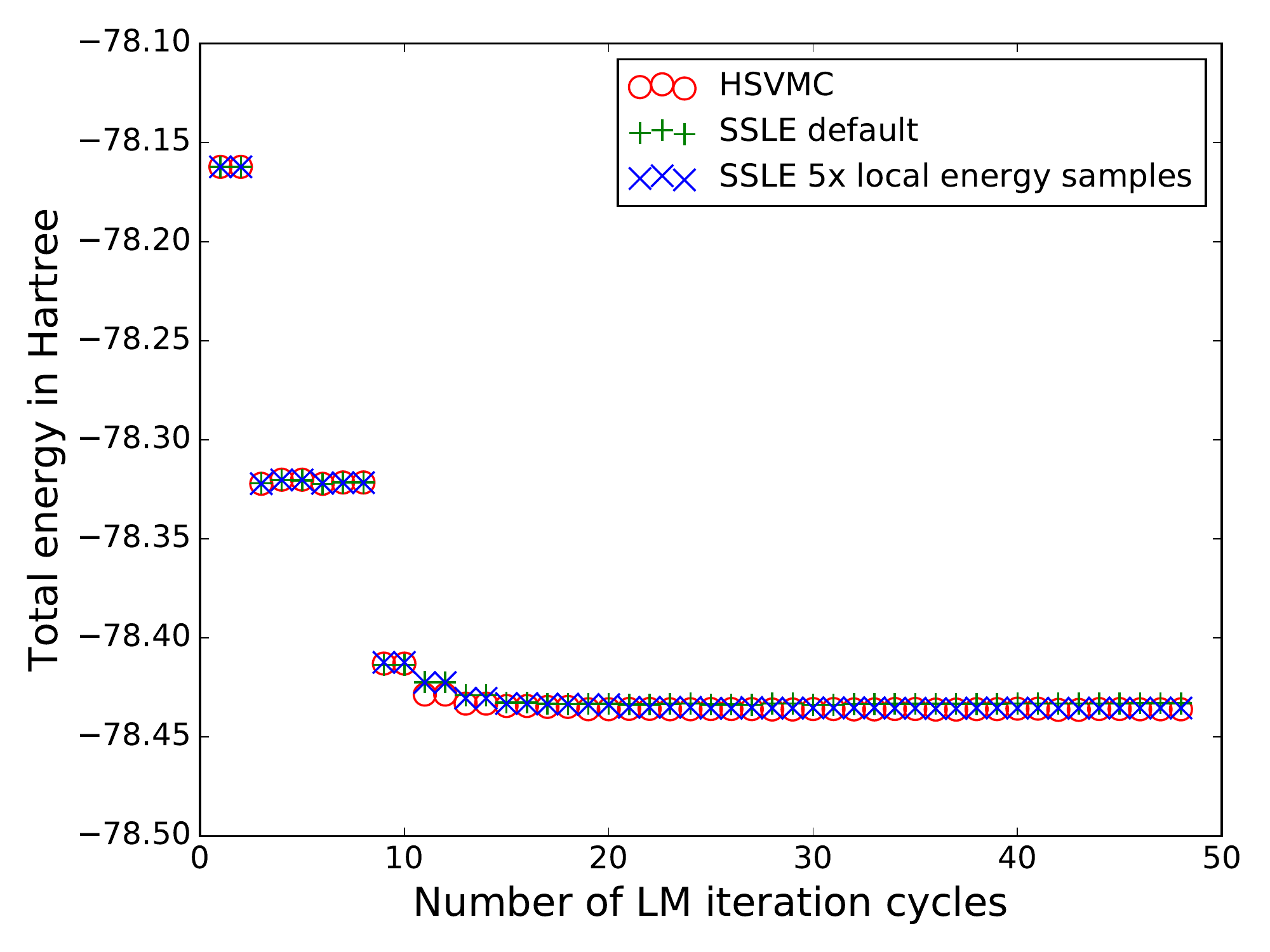}
  \caption{Total energy estimates of STO-3G ethane as the linear method optimizes the wave function.
           The Jastrow variables are fixed at zero for the first 8 iterations, after which they are optimized alongside the
           pairing matrix variables.
  }
  \label{fig:optimization_energy_plot}
\end{figure}

In addition to efficiency, however, one must also consider the fact that the derivatives and the optimization
methods that use them have different statistical properties than the energy itself.
For example, the nonlinear nature of the eigenvalue problem at the heart of the linear method can lead
statistical uncertainties in the local energy derivatives to lead to biased parameter updates.
\cite{Chris:2016:eom_hsvmc,schwarz2017projector}
Due to this concern, we have explicitly tested how the SSLE approach affects the wave function in the
context of a linear method optimization of the JAGP in ethane.
As seen in Figure \ref{fig:optimization_energy_plot}, the overall optimization is not greatly affected
by the use of the SSLE approach, although we do find that the additional uncertainty does slightly modify
the final optimized energy.
When using the two-electron index quadruplets sample length of $8n^2$ chosen in Section \ref{sec::ssle}, the final optimized
energy is 2 m$\mathrm{E}_{\mathrm{h}}$ higher than when the local energy and its derivatives are
evaluated fully deterministically.
Although it is difficult to tell if this poorer optimization is the result of a nonlinear bias or
simply a noisier but unbiased step direction, in either case the situation should improve if more
index quadruplet samples are used and the additional uncertainty from the SSLE is reduced.
Indeed, when we increase the index quadruplet sample size by a factor of 5, the optimized energy
is a more tolerable 0.4 m$\mathrm{E}_{\mathrm{h}}$ above the reference result.
Thus, we find that while the SSLE does not prevent successful wave function optimization,
the linear method appears to be more sensitive to the additional uncertainty created by the SSLE
approach than does the energy itself.
In future work on the SSLE, modifications to the optimization method that make it less sensitive
in this regard are clearly desirable, and it may be preferable to employ a descent optimizer
that is less prone to nonlinear bias difficulties and thus less noise sensitive.

\section{Conclusions}
\label{sec:conclusion}

We have presented a semi-stochastic local energy approach to Hilbert-space variational Monte Carlo that brings its cost
scaling with system size in line with the $O(n^4)$ overall cost scaling of real space variational Monte Carlo
for both Slater-Jastrow and Jastrow antisymmetric geminal power wave functions.
This scaling arises from an $O(n^3)$ cost per configuration due to the local energy evaluation and an additional
factor of $n$ due to the need to increase the configuration sample length in order to maintain a constant statistical
uncertainty as system size is increased.
We demonstrated that this scaling reduction does not come at the cost of a large prefactor, with a cost crossover
for energy evaluations occurring at around 50 valence electrons in our tests.
Furthermore, the resulting increase in statistical uncertainty is small and decays towards zero in the large system limit,
although it is somewhat more noticeable during a linear method optimization.
Given the wide range of ongoing activity in Hilbert-space variational Monte Carlo, especially in areas that connect
to machine learning, the ability to work with the ab initio molecular Hamiltonian at a lower cost scaling
should be broadly beneficial.

\section{Acknowledgements}

The authors acknowledge funding through the Early Career Research Program
of the Office of Science, Office of Basic Energy Sciences,
the U.S. Department of Energy, grant No.\ {DE-SC0017869}.
Computations were performed using the Berkeley Research Computing Savio cluster.

\bibliography{main}

\end{document}